\documentclass[12pt,aps,prb,preprint]{revtex4}   

\usepackage{amsmath}    
\usepackage{amsfonts}  
\usepackage{amssymb}
\usepackage{graphicx}   

\usepackage[latin1]{inputenc}     
\usepackage[T1]{fontenc}

\usepackage{float}   

\newtheorem{Theoreme}{THEOREM}
\newtheorem{Definition}{Definition}

\draft

\begin{document}

\title{The two limits of the Schrödinger equation in the semi-classical approximation:
discerned and non-discerned particles in classical mechanics}
\author{Michel Gondran}
\email{michel.gondran@polytechnique.org} \affiliation{University
Paris Dauphine, Lamsade, 75 016 Paris, France}
\author{Alexandre Gondran}
 \affiliation{Ecole Nationale de l'Aviation Civile, 31000 Toulouse, France}

\begin{abstract}
We study, in the semi-classical approximation, the convergence of
the quantum density and the quantum action, solutions to the
Madelung equations, when the Planck constant h tends to 0. We find
two different solutions which depend to the initial density . In
the first case where the initial quantum density is a classical
density $\rho_{0}(\mathbf{x})$, the quantum density and the
quantum action converge to a classical action and a classical
density which satisfy the statistical Hamilton-Jacobi equations.
These are the equations of a set of classical particles whose
initial positions are known only by the density
$\rho_{0}(\mathbf{x})$. In the second case where initial density
converges to a Dirac density, the density converges to the Dirac
function which corresponds to a unique classical trajectory.
Therefore we introduce into classical mechanics non-discerned
particles (case 1), which explain the Gibbs paradox, and discerned
particles (case 2). Finally we deduce a quantum mechanics
interpretation which depends on the initial conditions
(preparation), the Broglie-Bohm interpretation in the first case
and the Schrödinger interpretation in the second case.
\end{abstract}

 \maketitle

\section{Introduction}

Quantum mechanics came to exist in 1900 with the introduction of
the Planck constant h = $6,62 × 10^{-34} m^{2}$ kg /s. Its value
is very small and close to 0 in the units of classical mechanics.
But, $\hbar = h/2\pi$ =1 in the atomic units system (ua) used to
represent the atomic scale. What is the connection between quantum
mechanics and classical mechanics? We will particulary study the
semi-classical approximation when the Planck constant tends
towards 0. This approach usually presents two major difficulties:

- a mathematical difficulty to study the convergence of these
equations,

- a conceptual difficulty: the particles are considered to be
indistinguishable in quantum mechanics and distinguishable in
classical mechanics.

We will see how these two problems can be overcome.

\bigskip

Let us consider the wave function solution to the Schr\"odinger
equation $\Psi(\textbf{x},t)$:
\begin{equation}\label{eq:schrodinger1}
i\hslash \frac{\partial \Psi }{\partial t}=\mathcal{-}\frac{\hslash ^{2}}{2m}%
\triangle \Psi +V(\mathbf{x},t)\Psi
\end{equation}
\begin{equation}\label{eq:schrodinger2}
\Psi (\mathbf{x},0)=\Psi_{0}(\mathbf{x}).
\end{equation}
With the variable change $ \Psi
(\mathbf{x},t)=\sqrt{\rho^{\hbar}(\mathbf{x},t)} \exp(i
\frac{S^{\hbar}(\textbf{x},t)}{\hbar})$, the quantum density
$\rho^{\hbar}(\mathbf{x},t)$ and the quantum action
$S^{\hbar}(\textbf{x},t)$ depend on the parameter $\hbar$. The
Schrödinger equation can be decomposed into Madelung
equations~\cite{Madelung_1926} (1926) which correspond to:
\begin{equation}\label{eq:Madelung1}
\frac{\partial S^{\hbar}(\mathbf{x},t)}{\partial t}+\frac{1}{2m}
(\nabla S^{\hbar}(\mathbf{x},t))^2 +
V(\mathbf{x},t)-\frac{\hbar^2}{2m}\frac{\triangle
\sqrt{\rho^{\hbar}(\mathbf{x},t)}}{\sqrt{\rho^{\hbar}(\mathbf{x},t)}}=0
\end{equation}
\begin{equation}\label{eq:Madelung2}
\frac{\partial \rho^{\hbar}(\mathbf{x},t)}{\partial t}+ \nabla
\cdot (\rho^{\hbar}(\mathbf{x},t) \frac{\nabla
S^{\hbar}(\mathbf{x},t)}{m})=0  \qquad \forall (\mathbf{x},t)
\end{equation}
with initial conditions
\begin{equation}\label{eq:Madelung3}
\rho^{\hbar}(\mathbf{x},0)=\rho^{\hbar}_{0}(\mathbf{x}) \qquad and
\qquad S^{\hbar}(\mathbf{x},0)=S^{\hbar}_{0}(\mathbf{x}) .
\end{equation}

The Madelung equations are mathematically equivalent to the
Schrödinger equation if the functions
$\rho^{\hbar}_{0}(\mathbf{x})$ and $S^{\hbar}_{0}(\mathbf{x})$
exist and are smooth. It will be physically the case if $\Psi
^{0}(\mathbf{x})$ is a wave packet.

In this paper we study the convergence of the density
$\rho^{\hbar}(\mathbf{x},t)$ and the action
$S^{\hbar}(\textbf{x},t)$, solutions to the Madelung equations
when $\hbar$ tends to 0. Its convergence is subtle and remains a
difficult problem. We find, according to the assumptions on the
initial probability density $\rho_0^{\hbar}(\mathbf{x})$, two very
different cases of convergence \textit{due to a different
preparation of the particles}.

\begin{Definition}\label{defdensiteinitstat}- A quantum system
is prepared in the \textbf{statistical semi-classical case} if its
wave function satisfies the two following conditions

- its initial probability density $\rho^{\hbar}_{0}(\mathbf{x})$
and its initial action $S^{\hbar}_{0}(\mathbf{x})$ are regular
functions $\rho_{0}(\mathbf{x})$ and $S_{0}(\mathbf{x})$ not
depending on $\hbar$.

- its interaction with the potential field $V(\textbf{x},t)$ can
be described classically. The simplest case corresponds to
particles in a vacuum with only geometric constraints.
\end{Definition}

It is the case of a set of particles that are non-interacting and
prepared in the same way: a free particles beam in a linear
potential, an electronic or $C_{60}$ beam in the Young's slits
diffraction, an atomic beam in the Stern and Gerlach experiment.

\bigskip

\begin{Definition}\label{defdensiteinitponct}- A quantum system is prepared
in the \textbf{determinist semi-classical case} if its wave
function satisfies the two following conditions

- its initial probability density $\rho^{\hbar}_{0}(\mathbf{x})$
converges, when $\hbar\to 0$, to a Dirac distribution and its
initial action $S^{\hbar}_{0}(\mathbf{x})$ is a regular function
$S_{0}(\mathbf{x})$ not depending on $\hbar$.

- its interaction with the potential field $V(\textbf{x})$ can be
described classically.
\end{Definition}

This situation occurs when the wave packet corresponds to a
quasi-classical coherent state, introduced in 1926 by
Schr\"odinger~\cite{Schrodinger_26}. The field quantum theory and
the second quantification are built on these coherent
states~\cite{Glauber_65}. The existence for the hydrogen atom of a
localized wave packet whose motion is on the classical trajectory
(an old dream of Schr\"odinger's) was predicted in 1994 by
Bialynicki-Birula, Kalinski, Eberly, Buchleitner et
Delande~\cite{Bialynicki_1994, Delande_1995, Delande_2002}, and
discovered recently by Maeda and Gallagher~\cite{Gallagher} on
Rydberg atoms.

\bigskip

The separation of deterministic and statistical semi-classical
cases causes a strong reduction of the mathematical difficulties
of the convergence study.

In section 2, we show how, in the statistical semi-classical case,
the density $\rho^{\hbar}(\mathbf{x},t)$ and the action
$S^{\hbar}(\textbf{x},t)$, solutions to the Madelung equations,
converge, when the Planck constant $\hbar$ goes to zero, to a
classical density and a classical action which satisfy the
statistical Hamilton-Jacobi equations. These are the equations of
a set of classical particles whose initial positions are known
only by the density $\rho_{0}(\mathbf{x})$. Therefore we introduce
non-discerned particles into classical mechanics and the
Broglie-Bohm interpretation of the statistical semi-classical
case.

In section 3, we show how, in the determinist semi-classical case,
the density $\rho^{\hbar}(\mathbf{x},t)$ and the action
$S^{\hbar}(\textbf{x},t)$, solutions to the Madelung equations,
converge, when the Planck constant $\hbar$ goes to zero, to an
unique classical trajectory and an action which satisfy the
determinist Hamilton-Jacobi equations. Therefore we introduce
discerned particles into classical mechanics and the Schrödinger
interpretation of the determinist semi-classical case: the wave
function is then interpreted as the state of a single particle
similar to a soliton.

\section{Convergence in the statistical semi-classical case}

In the statistical semi-classical case, we have:

\begin{Theoreme} For particles in the statistical semi-classical case,
the probability density $\rho^{\hbar}(\textbf{x},t)$ and the
action $S^{\hbar}(\textbf{x},t)$, solutions to the Madelung
equations
(\ref{eq:Madelung1})(\ref{eq:Madelung2})(\ref{eq:Madelung3}),
converge, when $\hbar\to 0$, to the classical density
$\rho(\textbf{x},t)$ and the classical action $S(\textbf{x},t)$,
solutions to the statistical Hamilton-Jacobi equations:
\begin{equation}\label{eq:statHJ1}
\frac{\partial S\left(\textbf{x},t\right) }{\partial
t}+\frac{1}{2m}(\nabla S(\textbf{x},t) )^{2}+V(\textbf{x},t)=0
\end{equation}
\begin{equation}\label{eq:statHJ2}
S(\textbf{x},0)=S_{0}(\textbf{x}).
\end{equation}
\begin{equation}\label{eq:statHJ3}
\frac{\partial \mathcal{\rho }\left(\textbf{x},t\right) }{\partial
t}+ div \left( \rho \left( \textbf{x},t\right) \frac{\nabla
S\left( \textbf{x},t\right) }{m}\right) =0\text{ \ \ \ \ \ \ \
}\forall \left( \textbf{x},t\right)
\end{equation}
\begin{equation}\label{eq:statHJ4}
\rho(\mathbf{x},0)=\rho_{0}(\mathbf{x}) .
\end{equation}
\end{Theoreme}

We will demonstrate in the case where the wave function
$\Psi(\textbf{x},t)$ at time t is written as a function of the
initial wave function $\Psi_{0}(\textbf{x})$ by the Feynman
formula:

\begin{equation*}
\Psi(\textbf{x},t)= \int F(t,\hbar)
\exp(\frac{i}{\hbar}S_{cl}(\textbf{x},t;\textbf{x}_{0})
\Psi_{0}(\textbf{x}_{0})d\textbf{x}_0
\end{equation*}
where $F(t,\hbar)$ is an independent function of $\textbf{x}$ and
of $\textbf{x}_{0}$ and where
$S_{cl}(\textbf{x},t;\textbf{x}_{0})$ is the classical action
$min_{\textbf{u}(s)}\int_{0}^{t}L(\textbf{x}(s),\textbf{u}(s),s)
ds$, the minimum is taken over all trajectories $\textbf{x}(s)$
with velocity $\textbf{u}(s)$ from $\textbf{x}_0$ to $\textbf{x}$
between 0 and t.

Feynman\cite{Feynman_1965} (p. 58) shows that the general paths
integral formula is simplified in this form, especially when the
potential $ V(\textbf{x},t)$ is a quadratic function in
$\textbf{x}$.

\bigskip

Let us consider the statistical semi-classical case where
$\Psi_{0}(\textbf{x})= \sqrt{\rho_0(\mathbf{x})} \exp(i
\frac{S_0(\textbf{x})}{\hbar})$ with $ \rho_0(\mathbf{x})$ and $
S_0(\textbf{x})$ are non-dependent functions of $\hbar$. The wave
function is written
\begin{equation*}
\Psi(\textbf{x},t)= F(t,\hbar)\int\sqrt{\rho_0(\mathbf{x}_0)}
\exp(\frac{i}{\hbar}( S_0(\textbf{x}_0)+
S_{cl}(\textbf{x},t;\textbf{x}_{0})) d\textbf{x}_0.
\end{equation*}
The theorem of the stationary phase shows that, if $\hbar$ tends
towards 0, we have
\begin{equation*}
\Psi(\textbf{x},t)\sim \exp(\frac{i}{\hbar}min_{\textbf{x}_0}(
S_0(\textbf{x}_0)+ S_{cl}(\textbf{x},t;\textbf{x}_{0})).
\end{equation*}
that is to say that the quantum action $S^{h}(\textbf{x},t)$
converges to the function
\begin{equation}\label{eq:solHJminplus}
S(\textbf{x},t)=min_{\textbf{x}_0}( S_0(\textbf{x}_0)+
S_{cl}(\textbf{x},t;\textbf{x}_{0})).
\end{equation}
However, $S(\textbf{x},t) $ given by (\ref{eq:solHJminplus}) is
the solution to the Hamilton-Jacobi equation (\ref{eq:statHJ1})
with the initial condition (\ref{eq:statHJ2}). This is a
consequence of the principle of the least action and a fundamental
property of the minplus analysis we have
developed\cite{Gondran_1996,GondranMinoux_2008} following
Maslov\cite{MaslovSamborski_1992}.

Moreover, as the quantum density $\rho^{h}(\textbf{x},t)$ verifies
the continuity equation (\ref{eq:Madelung2}) of the Madelung
equations, we deduce, since $S^{h}(\textbf{x},t)$ tends towards
$S(\textbf{x},t)$, that $\rho^{h}(x,t)$ converges to the classical
density $\rho(\textbf{x},t)$, which satisfies the continuity
equation (\ref{eq:statHJ3}) of the statistical Hamilton-Jacobi
equations. We obtain both announced convergences.

\bigskip

This theorem will have major implications in classical and quantum
mechanics. The first one is to provide an interpretation of the
classical particles which satisfy the statistical Hamilton-Jacobi
equations.

\subsection{Non-discerned particules in classical mechanics }

The statistical Hamilton-Jacobi equations correspond to a set of
independent classical particles, in a potential field
$V(\mathbf{x},t)$, and for which we only know at the initial time
the probability density $\rho _{0}\left( \mathbf{x}\right) $ and
the velocity $\mathbf{v(x)}=\frac{\nabla S_{0}(\mathbf{x},t)}{m}$.

Let us consider N particles that satisfy the statistical
Hamilton-Jacobi equations. We propose the following definition:

\begin{Definition}\label{defNparticulesindiscernedenmc}- N indentical particles,
prepared in the same way, with the same initial density $ \rho
_{0}\left( \textbf{x}\right)$, the same initial action
$S_0(\textbf{x})$, and evolving in the same potential
$V(\textbf{x},t)$ are called \textbf{non-discerned}.
\end{Definition}
We refer to these particles as non-discerned and not as
indistinguishable because, if their initial positions are known,
their trajectories will be known as well. Nevertheless, when one
counts them, they will have the same properties as the
indistinguishable ones. Thus, if the initial density $\rho
_{0}\left( \textbf{x}\right)$ is given, and one randomly chooses
$N$ particles, the N! permutations are strictly equivalent and do
not correspond to the same configuration as for indistinguishable
particles. This means that if $X$ is the coordinate space of a
non-discerned particle, the true configuration space of $N$
non-discerned particles is not $X^N$ but rather $ X^N / S_N $
where $S_N$ is the symmetric group.

The introduction of these non-discerned particles into classical
mechanics solves the conceptual difficulty announced in the
introduction; indiscernibility also exists in classical mechanics.
These non-discerned particles in classical mechanics also give a
simple solution to the Gibbs paradox. This view is not new: it
features particularly in Landé\cite{Lande_1965} in 1965, Leinaas
et Myrheim~\cite{Leinaas_1976} in 1977 and more recently in
Greiner~\cite{Greiner_1999} in his book "Thermodynamics and
statistical mechanics".

\subsection{Quantum trajectories of de Broglie-Bohm}

In the statistical semi-classical case, the Madelung equations
converge to statistical Hamilton-Jacobi equations. The
uncertainity about the position of a quantum particle corresponds
in this case to an uncertainity about the position of a classical
particle, only whose initial density has been defined. \textbf{In
classical mechanics, this uncertainity is removed by giving the
initial position of the particle. It would not be logical not to
do the same in quantum mechanics.}

\bigskip
We assume that for \textit{the statistical semi-classical case}, a
quantum particle is not well described by its wave function. One
needs therefore to add its initial position and it becomes natural
to introduce the so-called de Broglie-Bohm trajectories. In this
interpretation, its velocity is given by
\cite{deBroglie_1927,Bohm_52}:
\begin{equation}\label{eq:vitessequantique}
\textbf{v}^{\hbar}(\textbf{x},t) = \frac{1}{m}\nabla
S^{\hbar}(\textbf{x},t).
\end{equation}

We have the classical property: if a system of particles with
initial density $\rho_{0}(\mathbf{x})$ follows de Broglie-Bohm
trajectories defined by the velocity field
$\textbf{v}^{\hbar}(\textbf{x},t)$, then the probability density
of those particles at time $t$ is equal to
$\rho^\hbar(\textbf{x},t)$, the square of the wave function. Using
this velocity, the Heisenberg inequalities correspond to a
dispersion relation position and velocity between the different
non-discerned particles. This shows that, in the statistical
semi-classical case, the Broglie-Bohm interpretation reproduces
the predictions of standard quantum mechanics.

Therefore, when $\hbar\to 0$, we deduce that
$\textbf{v}^{\hbar}(\textbf{x},t)$ given from equation
(\ref{eq:vitessequantique}) converges to the classical velocity
$\textbf{v}(\textbf{x},t) = \frac{1}{m}\nabla S(\textbf{x},t)$ and
we obtain:

\begin{Theoreme} For particles in the statistical semi-classical
case,when $\hbar\to 0$, the de Broglie-Bohm trajectoires converge
to the classical ones.
\end{Theoreme}

Figure 1 shows a simulation of the Broglie-Bohm's trajectories in
the Young slits experiment of Jönsson\cite{Jonsson_1961} where an
electron gun emits electrons one by one through a hole with a
radius of a few micrometers. We are in a statistical
semi-classical case where the electrons, prepared similarly, are
represented by the same initial wave function, but not by the same
initial position. In the simulation, these initial positions are
randomly selected in the initial wave packet. We have represented
100 possible quantum trajectories through one of two slits: we do
not show the trajectories of the electron when it is stopped by
the first screen.

\begin{figure}[H]
\begin{center}
\includegraphics[width=0.6\linewidth]{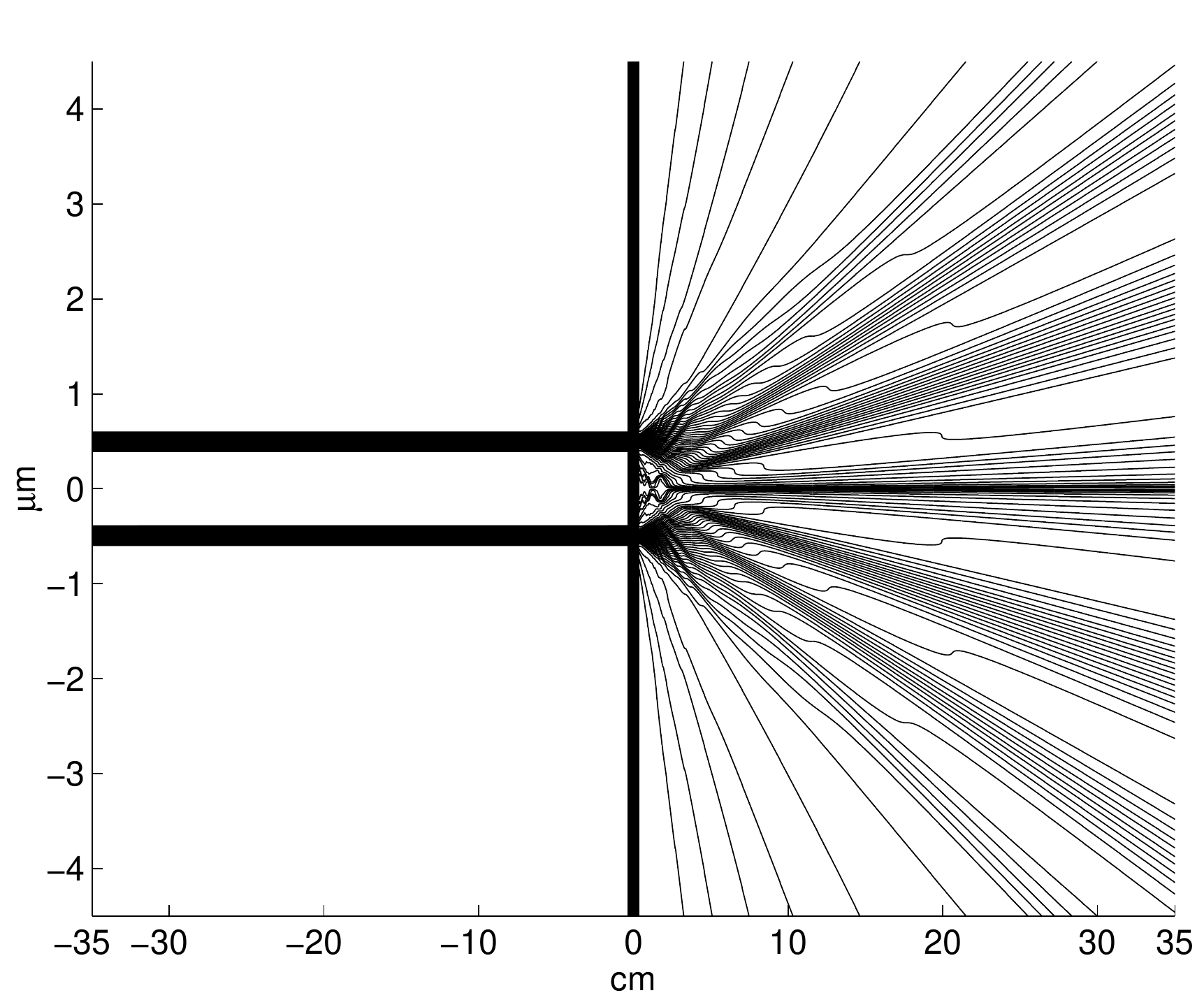}
\caption{\label{fig:trajectoiresBohmhydro} 100 Broglie-Bohm's
trajectories through one of two slits.}
\end{center}
\end{figure}

Figure 2 shows the 100 previous trajectories when the Planck
constant is divided by 10, 100, 1000 and 10000 respectively. We
obtain, when h tends to 0, the convergence of quantum trajectories
to classic trajectories.
\begin{figure}[H]
\begin{center}
\includegraphics[width=0.8\linewidth]{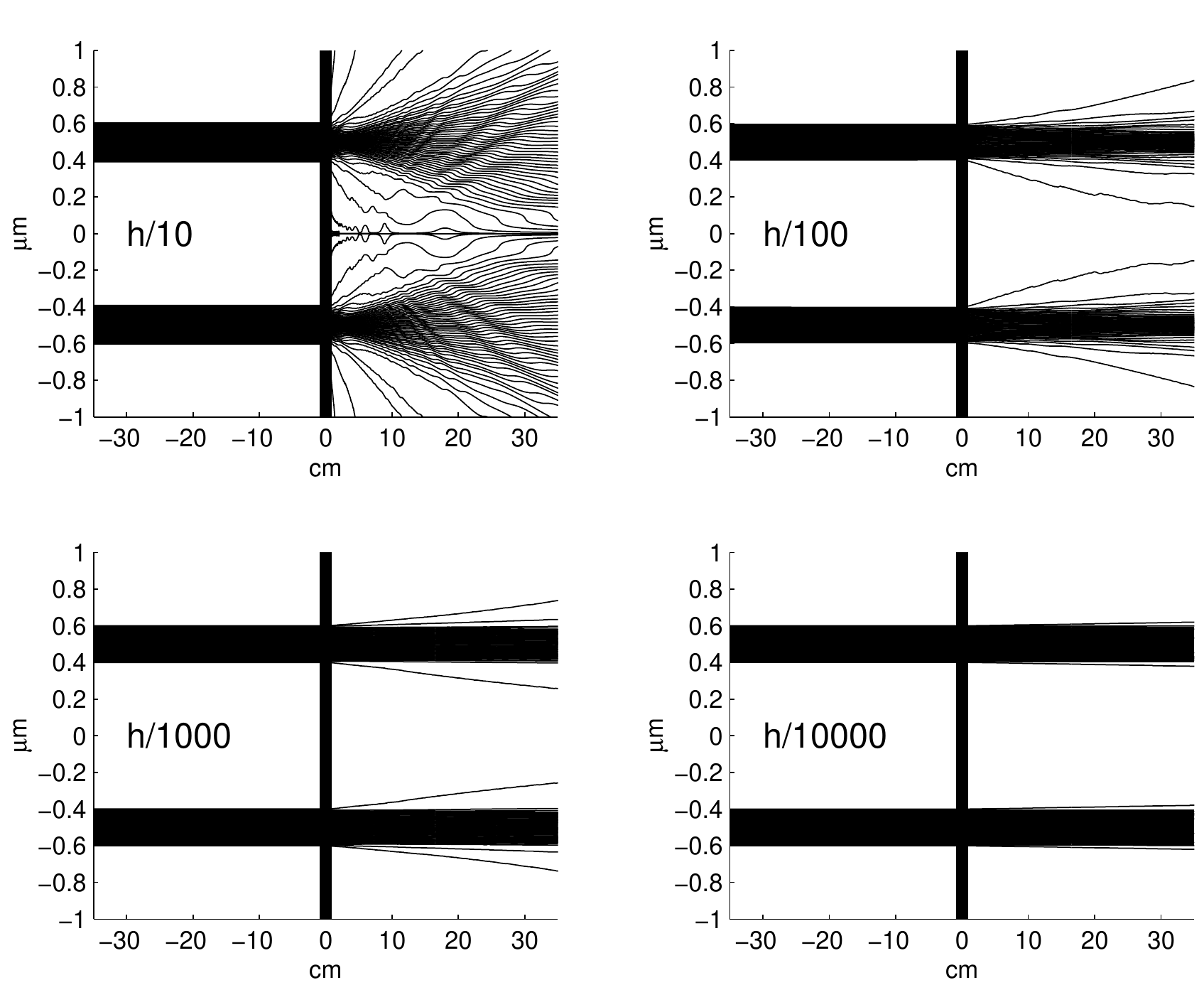}
\caption{\label{fig:trajectoiresBohmhydro2} Zoom of the 100
previous trajectories when the Planck constant is divided by 10,
100, 1000 and 10000 respectively.}
\end{center}
\end{figure}

\section{Convergence in the determinist semi-classical case}

The convergence study of the determinist semi-classical case is
mathematically very difficult. We only study the example of a
coherent state where an explicit calculation is possible. This
example will enable to understand the convergence in the
determinist semi-classical case and the difference with the
statistical semi-classical case.

For the two dimensional harmonic oscillator,
$V(\textbf{x})=\frac{1}{2}m \omega^{2}\textbf{x}^{2}$, coherent
states are built~\cite{CohenTannoudji_1977} from the initial wave
function $\Psi_{0}(\textbf{x})$ which corresponds to the density
and initial action :
\begin{equation}\label{eq:densiteactioninith}
\rho^{\hbar}_{0}(\mathbf{x})= ( 2\pi \sigma _{\hbar}^{2}) ^{-1}
e^{-\frac{( \textbf{x}-\textbf{x}_{0}) ^{2}}{2\sigma
_{\hbar}^{2}}} \qquad and \qquad
S_{0}(\mathbf{x})=S^{\hbar}_{0}(\mathbf{x})= m \textbf{v}_{0}\cdot
\textbf{x}
\end{equation}
with $ \sigma_\hbar=\sqrt{\frac{\hbar}{2 m \omega}}$. Here,
$\textbf{v}_0$ and $\textbf{x}_0$ are still constant
vectors and independant from $\hbar$, but $\sigma_\hbar$ will tend to $0$ as $\hbar$.

For this harmonic oscillator, the density
$\rho^{\hbar}(\textbf{x},t)$ and the action
$S^{\hbar}(\textbf{x},t)$,solutions to the Madelung equations
(\ref{eq:Madelung1})(\ref{eq:Madelung2})(\ref{eq:Madelung3}) with
initial conditions (\ref{eq:densiteactioninith}), are equal to
~\cite{CohenTannoudji_1977}:
\begin{equation}
\rho^{\hbar}(\textbf{x},t)=\left( 2\pi \sigma_{\hbar} ^{2} \right)
^{-1}e^{- \frac{( \textbf{x}-\xi(t)) ^{2}}{2\sigma_{\hbar} ^{2}
}}~~~~and~~~~S^{\hbar}(\textbf{x},t)=  + m \frac{d\xi (t)}{dt}\cdot
\textbf{x} + g(t) - \hbar\omega t
\end{equation}
where $\xi(t)$ is the trajectory of a classical particle evolving
in the potential $V(\textbf{x})=\frac{1}{2} m \omega^{2}
\textbf{x}^2 $, with $\textbf{x}_0$ and $\textbf{v}_0$ as initial
position and velocity and $g(t)=\int _0 ^t ( -\frac{1}{2} m
(\frac{d\xi (s)}{ds})^{2} + \frac{1}{2} m \omega^{2} \xi(s)^2)
ds$. Because we have $2 V(\xi(s))=m \frac{d^2 \xi(s)}{ds^2} \cdot
\xi(s) $, it yields the following theorem:

\begin{Theoreme}\label{t-convergenceparticulediscerne}- When $\hbar\to 0$,
the density $\rho^{\hbar}(\textbf{x},t)$ and the action
$S^{\hbar}(\textbf{x},t)$ converge to
\begin{equation}
\rho(\textbf{x},t)=\delta( \textbf{x}- \xi(t)) ~~ and ~~
S(\textbf{x},t)= m \frac{d\xi (t)}{dt}\cdot\textbf{x} + g(t)
\end{equation}
where $S(\textbf{x},t)$ and the trajectory $\xi(t)$ are solutions
to the determinist Hamilton-Jacobi equations:
\begin{equation}\label{eq:statHJponctuelle1b}
\frac{\partial S\left(\textbf{x},t\right) }{\partial
t}|_{\textbf{x}=\xi(t)}+\frac{1}{2m}(\nabla S(\textbf{x},t)
)^{2}|_{\textbf{x}=\xi(t)}+V(\textbf{x})|_{\textbf{x}=\xi(t)}=0\text{
\ \ \ \ \ \ \ }\forall t \in \mathbb{R}^{+}
\end{equation}
\begin{equation}\label{eq:statHJponctuelle1c}
\frac{d\xi(t)}{dt}=\frac{\nabla S(\xi(t),t)}{m}\text{ \ \ \ \ \ \
\ }\forall t \in \mathbb{R}^{+}
\end{equation}
\begin{equation}\label{eq:statHJponctuelle1d}
S(\textbf{x},0)= m \textbf{v}_0 \cdot \textbf{x}
~~~and~~~\xi(0)=\textbf{x}_0.
\end{equation}
\end{Theoreme}
Therefore, the kinematic of the wave packet converges to the
single harmonic oscillator one described by $\xi(t)$. Because this
classical particle is completely defined by its initial conditions
$\textbf{x}_0$ and $\textbf{v}_0$, it can be considered as a
discerned particle.

When $\hbar\to 0$, the "quantum potential"
$Q^{\hbar}(\textbf{x},t)=-\frac{\hbar^2}{2m}\frac{\triangle
\sqrt{\rho}}{\sqrt{\rho}}= \hbar \omega - \frac{1}{2} m \omega^{2}
(\textbf{x} -\xi(t))^{2}$ tends to $Q(\textbf{x},t)= - \frac{1}{2}
m \omega^{2} (\textbf{x} -\xi(t))^{2}$. It is then zero on the
trajectory ($\textbf{x}=\xi(t)$).

It is then possible to consider, unlike the statistical
semi-classical case, that the wave function can be viewed as a
single quantum particle. The \textit{determinist semi-classical
case} is in agreement with the Copenhagen interpretation of the
wave function which contains all the information on the particle.

\subsection{Interpretation for the determinist semi-classical case}

In the \textit{determinist semi-classical case}, the Broglie-Bohm
interpretation is not relevant mathematically unlike the
statistical semi-classical case. Other assumptions are possible. A
natural interpretation is the one proposed by Schr\"odinger
~\cite{Schrodinger_26} in 1926 for the coherent states of the
harmonic oscillator. In \textbf{Schr\"odinger interpretation}, the
quantum particle in the determinist semi-classical case is a
spatially extended particle, represented by a wave packet whose
center follows the classical trajectory. In this interpretation,
the velocity in each point of the wave function is given
by\cite{Bohm_93, Holland_93, Holland_99, GondranMA}:
\begin{equation}\label{eq:vitessequantiqueavecspin}
\textbf{v}^{\hbar}(\textbf{x},t) = \frac{1}{m}\nabla
S^{\hbar}(\textbf{x},t) + \frac{\hbar}{2 m}\nabla
\ln\rho^{\hbar}(\textbf{x},t) \times \textbf{k},
\end{equation}
where $\textbf{k}$ is the unit vector parallel to the particle
spin vector. This spin current $\frac{\hbar}{2 m}\nabla
\rho^{\hbar}(\textbf{x},t) \times \textbf{k}$ corresponds to
Gordon's current when one changes from Dirac's equation to Pauli's
equation and after that to Schrodinger's
equation\cite{Holland_99}. This current is important because it
allows to revisit quantum mechanics at small scales, in particular
Compton's wavelength as in the Foldy and Wouthuysen transformation
\cite{Foldy}. Using this velocity, the Heisenberg inequalities
correspond to a dispersion relation position and velocity between
the different points of the extended particle.

For the coherent states of the harmonic oscillator in two
dimensions, the velocity field (\ref{eq:vitessequantiqueavecspin})
is equal to:
\begin{equation}\label{eq:eqvitessemqosc}
\textbf{v}^{\hbar}(\mathbf{x},t)=\textbf{v}(t) + \Omega \times
(\textbf{x} -\xi(t))
\end{equation}
with $\Omega = \omega \textbf{k}$. They behave as extended
particles which have the same evolution as spinning particles in
two dimensions. But this can not be generalized easily in three
dimensions. It seems that it is not possible to consider in three
dimensions the particle as a solid in motion. This is the main
difficulty in the Schrödinger interpretation: does the particle
exist within the wave packet? We think that this reality can only
be defined at the scale where the Schrödinger equation is the
effective equation. Some solutions are nevertheless possible at
smaller scales~\cite{Gondran_2001,Gondran_2004}, where the quantum
particle is not represented by a point but is a sort of elastic
string whose gravity center follows the classical trajectory
$\xi(t)$.

\subsection{Interpretation for the non semi-classical case}

The Broglie-Bohm and Schrödinger interpretations correspond to the
semi-classical approximation. But there exist situations where the
semi-classical approximation is not valid. It is in particular the
case of state transitions for a hydrogen atom. Indeed, since
Delmelt'experiment~\cite{Delmelt_1986} in 1986, the physical
reality of individual quantum jumps has been fully validated. The
semi-classical approximation, where the interaction with the
potential field can be described classically, is not possible
anymore and it is necessary to electromagnetic field quantization
since the exchanges are done photon by photon.

In this situation, the Schrödinger equation cannot give a
deterministic interpretation and the statistical Born
interpretation is the only valid one.

These three interpretations are not new, as Einstein points out in
one of his last articles (1953), "\textit{Elementary reflexion
concerning the quantum mechanics foundation}" in a homage to Max
Born:

"\textit{The fact that the Schr\"odinger equation associated to
the Born interpretation does not lead to a description of the
"real states" of an individual system, naturally incites one to
find a theory that is not subject to this limitation.} \textit{Up
to now, the two attempts have in common that they conserve the
Schr\"odinger equation and abandon the Born interpretation.}
\textit{\textbf{The first one}, which marks \textbf{de Broglie's
comeback}, was continued by Bohm.}... \textit{\textbf{The second
one}, which aimed to get \textbf{a "real description" of an
individual system} and which might be based on the Schr\"odinger
equation is very late and is from Schr\"odinger himself. The
general idea is briefly the following : \textbf{the function
$\psi$ represents in itself the reality} and it is not necessary
to add it to Born's statistical interpretation.}[...] \textit{From
previous considerations, it results that the only acceptable
interpretation of the Schr\"odinger equation is the statistical
interpretation given by Born. Nevertheless, this interpretation
doesn't give the "real description" of an individual system, it
just gives statistical statements of entire systems.}"

What is new is to consider that these interpretations depend on
the preparation of the particles.

Thus, it is because de Broglie and Schr\"odinger keep the
Schrödinger equation that Einstein, who considers it as
fundamentaly statistical, rejected each of their interpretations.

Einstein thought that it was not possible to obtain an individual
deterministic behavior from the Schrödinger equation. It is the
same for Heisenberg who developped matrix mechanics and the second
quantization from this example.

This doesn't mean that one has to renounce determinism and
realism, but rather that Schr\"odinger's statistical wave function
does not permit, in that case, to obtain an individual behavior.

\bigskip

\section{Conclusion}

The study of the convergence of the Madelung equations when $\hbar
\to 0$, has encouraged us to introduce the concept of
non-discerned and discerned particles in classical mechanics and
has given us the three following results:

- In \textit{\textbf{the statistical semi-classical case}} the
quantum particles converge to classical non-discerned ones
verifying the statistical Hamilton-Jacobi equations. The wave
function is not sufficient to represent the quantum particles. One
needs to add it to the initial positions, as for classical
particles, in order to describe them completely. Then,
\textit{\textbf{the Broglie-Bohm interpretation is relevant}}.

- In \textit{\textbf{the determinist semi-classical case}} the
quantum particles converge to classical discerned ones verifying
the determinist Hamilton-Jacobi equations. \textit{\textbf{The
Broglie-Bohm interpretation is not relevant}} because the wave
function is sufficient to represent the particles as in the
Copenhagen interpretation. However, one can make a realistic and
deterministic assumption such as the \textbf{Schr\"odinger
interpretation}.

- In the case where \textit{\textbf{the semi-classical
approximation is not valid anymore}}, as in the transition states
in the hydrogen atom, the two interpretations are wrong as claimed
by Heisenberg. Consequently, \textbf{Born's statistical
interpretation} is the only possible interpretation of the
Schrödinger equation. This doesn't mean that it is necessary to
abandon determinism and realism, but rather that the Schr\"odinger
wave function doesn't allow, in that case, to obtain an individual
behavior of a particle. An individual interpretation needs to use
creation and annihilation operators of the quantum Field Theory.

Therefore, as Einstein said, the situation is much more complex
than de Broglie and Bohm thought.

Each founding father of quantum mechanics held a piece of the
truth; but the overgeneralization of their different truths has
led to incompatible interpretations!

\end{document}